# Anisotropic $g$-Factor and Spin-Orbit Field in a Ge Hut Wire Double Quantum Dot

Ting Zhang,[1,2, #] He Liu,[1,2, #] Fei Gao,[3, #] Gang Xu,[1,2] Ke Wang,[1,2] Xin Zhang[1,2], Gang Cao,[1,2] Ting Wang,[3] Jianjun Zhang,[3] Xuedong Hu[4], Hai-Ou Li,[1,2,*] and Guo-Ping Guo[1,2,5,*]

[1] *CAS Key Laboratory of Quantum Information, University of Science and Technology of China, Hefei, Anhui 230026, China*

[2] *CAS Center for Excellence and Synergetic Innovation Center in Quantum Information and Quantum Physics, University of Science and Technology of China, Hefei, Anhui 230026, China*

[3] *Institute of Physics and CAS Center for Excellence in Topological Quantum Computation, Chinese Academy of Sciences, Beijing 100190, China*

[4] *Department of Physics, University at Buffalo, SUNY, Buffalo, New York 14260, USA*

[5] *Origin Quantum Computing Company Limited, Hefei, Anhui 230026, China*

[#]These authors contributed equally to this work

\* Corresponding author. Emails: haiouli@ustc.edu.cn (H.-O. L.); gpguo@ustc.edu.cn (G.-P.G.).

## Abstract

Holes in nanowires have drawn significant attention in recent years because of the strong spin-orbit interaction, which plays an important role in constructing Majorana zero modes and manipulating spin-orbit qubits. Here, from the strongly anisotropic leakage current in the spin blockade regime for a double dot, we extract the full $g$-tensor and find that the spin-orbit field is in plane with an azimuthal angle of 59° to the axis of the nanowire. The direction of the spin-orbit field indicates a strong spin-orbit interaction along the nanowire, which may have originated from the interface inversion asymmetry in Ge hut wires. We also demonstrate two different spin relaxation mechanisms for the holes in the Ge hut wire double dot: spin-flip cotunneling to the leads, and spin-orbit interaction within the double dot. These results help establish feasibility of a Ge-based quantum processor.

# Main text

The spin-orbit interaction (SOI) can be considered as a coupling between the momentum of a carrier and its spin. When an electron or hole moves in an electric field, it experiences an effective magnetic field that couples to its spin degree of freedom, even in the absence of an external magnetic field[1-3]. Recently, studies on the SOI have received widespread attention because of its fundamental role in both classical spintronics and the quantum coherent manipulation of a spin qubit[4-19], and more recently in the search for Majorana fermions[20-25]. The SOI can be exploited for all-electrical manipulation of a spin-orbit qubit using electric dipole spin resonance (EDSR) techniques[4-19,26-31]. Without generating a local ac magnetic field or making use of an external source of spin-electric coupling (such as micromagnets), the device design is simplified by using a local ac electric field. In addition, as proposed by theory and confirmed by recent experiments, a strong SOI is required to create Majorana fermions in semiconductor-superconductor heterostructures[20-25,32,33]. Thus, the potential applications of the SOI for quantum information processing will be particularly relevant in the future, especially with respect to fast operation speed[31], scalability[34], and topological quantum computation[35].

At the mean field level, SOI can be characterized with an effective magnetic field $\vec{B}_{SO}$. Here, the amplitude of $\vec{B}_{SO}$ reflects the strength of the SOI, which is related to the Rabi frequency in EDSR[31], and the direction of $\vec{B}_{SO}$ determines the suitable geometries in experimental setups. For example, the special geometry $\vec{B} \perp \vec{B}_{SO}$ is assumed in proposals for realizing Majorana fermions in nanowires with proximity-induced superconductivity[32,33]. Furthermore, EDSR is efficient when the externally applied magnetic field $\vec{B}$ is perpendicular to $\vec{B}_{SO}$[36-38]. However, under such conditions, the degree of mixing between the singlet and triplet states in a double quantum dot also reaches a maximum, which is expected to be proportional to $|\vec{B}_{SO} \times (\vec{B}/B)|$[39-41]. The transitions between T(1,1) and S(0,2) induced by the SOI hybridization have an adverse effect on the fidelity of the initialization and readout of spin qubits[39,42-44], and could speed up decoherence[45-52].

Holes in Ge hut wires (HWs)[53] are recognized as a prospective platform for quantum computation because of their favourable properties. The extremely strong "direct Rashba spin-orbit interaction" (DRSOI)[54-56] in Ge HWs enables a fast Rabi frequency of 540 MHz[18], which is the record manipulation speed for semiconductor spin qubits up to now. Furthermore, the growth of site-controlled Ge HWs has been demonstrated recently[34], which opens a potential path towards large-scale multi-qubit device fabrication. To establish the feasibility of hole-based quantum processors, and to possibly observe novel phenomena such as the Majorana zero mode in Ge HWs, it is imperative to investigate the spin-orbit field (SOF), in terms of both its magnitude and its direction.

To date, by analysing the anisotropic effects of the SOI, the SOF direction has been determined in various semiconducting nanostructures such as hole quantum dots (QDs) in the GaAs/AlGaAs heterostructure[43], silicon MOS[57,58], and electron QDs made from InAs nanowires[59]. However, experimental determination of the SOF direction, which is related to the DRSOI in hole nanowires, remains lacking.

In this letter, we extract the anisotropic $g$-tensor and determine the SOF direction in an electrostatically defined double quantum dot (DQD) in a Ge hut wire by analysing the behavior of leakage in the Pauli spin blockade (PSB) region. First, we measure the dependence of leakage current on the magnetic field along different directions, which can be attributed to the relaxation mechanisms of spin-flip cotunneling to the reservoirs and SOI within the DQD. The effective $g$-factors in-plane can be determined by fitting to the current peak induced by spin-flip cotunneling. Then, the magnetic field is fixed at a value dominated by the SOI. We study the anisotropic effects of the leakage current and extract the direction of $\vec{B}_{SO}$, which is in-plane and at a 59° angle to the nanowire, indicating the existence of other SOI mechanisms in addition to the DRSOI.

The DQD used in the experiment is fabricated from Ge HWs that are monolithically grown on a Si substrate[53]. Figure 1a shows a false-coloured scanning electron microscopy (SEM) image of the device we studied with five electrodes above a 1-μm-long nanowire. After removing oxides with buffered hydrofluoric acid, 30-nm-thick palladium contacts are patterned with electron beam lithography. A 20-nm aluminium oxide layer is then deposited as a gate dielectric using atomic layer deposition. Finally, three 30-nm-wide top gates consisting of 3-nm titanium and 25-nm palladium are fabricated between contact pads, spaced at a 40 nm pitch. A 3D schematic of the device is depicted in Figure 1b. By applying voltages to three top gates to create a confinement potential, a DQD is defined along the $y$-axis (parallel to the Ge HW), as shown in Figure 1c. Here, the experiment is carried out in a dilution refrigerator at a base temperature of about 15 mK.

Figure 1d shows the charge stability diagram of the DQD. We measure the source-drain current $I_{SD}$ as a function of gate voltages $V_L$ and $V_R$ at a bias voltage $V_{SD} = +2.5$ mV. Note that only when the energy levels corresponding to transitions of the two QDs both enter the bias window can the current through the DQD be detected, leading to an array of characteristic bias triangles[51,60]. We investigate the bias triangle denoted by a red circle at the transition between the (1,1) and (0,2) states in Figure 1d. Here, the number in parentheses describes the effective hole occupation of each quantum dot, while the true hole occupation is (m+1, n+1) and (m, n+2). When the total hole spin of each dot is zero in the (m,n) state, the spin-related transition between (m+1, n+1) and (m, n+2) states can be described with (1,1) and (0,2) states for simplicity. Figures 2a and 2b display zoom-ins of this triangle for positive and negative source-drain biases. The corresponding line traces along the detuning energy $\varepsilon$ (green arrow) are plotted in Figure 2c. In comparison, we observe a strong suppression of current at $V_{SD} = +2.5$ mV, which is attributed to the forbidden transition from T(1,1) to S(1,1) due to spin conservation during hole tunneling (Figure 2d), the characteristic signature of PSB[61-65]. For $V_{SD} = -2.5$ mV, a large current is measured in the whole triangle.

The spin blockade can be lifted by spin relaxation mechanisms such as hyperfine interaction[62,67,68], spin-flip cotunneling[66,69-71], a $g$-factor difference in the DQD and the SOI[39,44,67,69,72]. The dominant spin relaxation mechanism can be determined based on the leakage current in the PSB region. Figure 3 shows a measurement of the leakage spectroscopy as a function of $\varepsilon$ and $\vec{B}$ in different directions. Here, $\varepsilon$ can be tuned by changing the voltages of gates L and R as described in Figure 2a (green arrow), and the direction of $\vec{B}$ is depicted on the top panels of Figure 3. Figure 3a shows the leakage spectroscopy when the magnetic field is along the $z$ direction. The leakage current $I_{SD}$ in the PSB region increases monotonically as $B_z$ increases. A line cut along $\varepsilon = 0$ is plotted (red circles) in Figure 3d. The PSB is lifted at a finitely large magnetic field, and the current profile shows a broad dip, reflecting the presence of a spin-nonconserving transport mechanism. In Ge HWs, the strong SOI can hybridize the states of T(1,1) and S(0,2) at a finite magnetic field and enable a spin-flip transition between them to lift the PSB[39], leading to the leakage spectroscopy shown in Figure 3a.

However, when the direction of magnetic field is rotated perpendicularly and parallelly to the nanowire in plane, we find a different field-dependent behaviour of the leakage current, as shown in Figures 3b and 3c. From the line cut at $\varepsilon = 0$ (Figures 3e and 3f), we find the leakage curve is composed of a zero-field peak and a dip at a large magnetic field, which is induced by two different spin relaxation mechanisms. We attribute the broad leakage dip to the SOI in our system and find that the current increment induced by the SOI with an in-plane magnetic field is much smaller than that with an out-of-plane magnetic field. Although no saturation of leakage current is observed as before[43], we find that the dip widths are dramatically different in different directions. This phenomenon can be explained with the anisotropy of the $g$-factor in our hole system and the change in the angle between $\vec{B}$ and $\vec{B}_{SO}$[43], which will be discussed in detail below.

Furthermore, because the width of the peak is approximately 200 mT, we rule out the hyperfine interaction[62,68] and consider spin-flip cotunneling[66,69] to be the dominant mechanism for lifting of the PSB around zero field. When a triplet T(1,1) is formed in a DQD, one of the spins can flip through inelastic cotunneling between the dot and its nearest lead, giving rise to an enhanced leakage current. The spin-flip cotunneling rate will decrease exponentially with the increase of Zeeman energy[66]. Therefore, the leakage current induced by spin-flip cotunneling has a maximum at zero field and drops to zero at a large magnetic field, leading to a peak of leakage current

The leakage current $I_{\text{leak}}$ induced by SOI and spin-flip cotunneling can be expressed by[39,69]

$$I_{\text{leak}} = \frac{4ecg^*\mu_B B}{3\sinh\frac{g^*\mu_B B}{k_B T}} + I_{\text{SO}}^0 \frac{B^2}{B^2+B_C^2} + I_B, \quad (1)$$

The first term in the formula describes the contribution from the spin-flip cotunneling, with $c = \frac{h}{\pi}[\{\Gamma_l/\Delta\}^2 + \{\Gamma_r/(\Delta - 2U_M - 2eV_{\text{SD}})\}^2]$, where $e$ is the hole charge, $g^*$ is the effective $g$-factor, $\mu_B$ is the Bohr magneton, $h$ is Plank's constant, $k_B$ is the Boltzmann constant, T is the hole temperature, $\Gamma_{l,(r)}$ is the tunnel coupling with the left (right) reservoir, $\Delta$ is the depth of the two-hole level, and $U_M$ is the mutual charging energy[71]. The second term in the formula represents the SOI induced leakage current, which exhibits a Lorentzian-shaped dip with a width $B_C$[39]. And the residual leakage current $I_B$ results from other spin relaxation mechanisms, such as the difference of $g$-factor between two dots.

The black solid lines in Figures 3d-3f show the fitting to experimental data of the leakage current at $\varepsilon = 0$ with eq (1). Obviously, the experimental data are in good agreement with the theory. From the fitting to the current peak induced by spin-flip cotunneling, we can extract the values of the effective $g$-factor $g_x^* = g_y^* = 1.2 \pm 0.2$ with a hole temperature T = 40 mK. The obtained $g$-factors for a magnetic field applied in-plane are consistent with those previously extracted for a single QD[73]. By fitting the data in Figures 3d-3f in three different directions and focusing on the SOI part of eq (1), we can extract the values of $B_C$, with $B_C^x = 9$ T, $B_C^y = 6.5$ T and $B_C^z = 1.7$ T. The values of $B_C$ are so large that they exceed the limit of the magnetic field we can reach, which is why we cannot observe saturation of the leakage current induced by the SOI at a finite field. Here, $B_C$ is related not only to the effective $g$-factor but also to the angle between $\hat{g} \cdot \vec{B}$ and $\vec{B}_{\text{SO}}$, and it can be expressed as[43,57]

$$B_{\text{C}} = \frac{B_{\text{C}}^0}{g^* \cdot \sin\alpha}, \tag{2}$$

where $B_{\text{C}}^0$ is a constant, $g^*$ is the effective $g$-factor along the direction of magnetic field and $\alpha$ is the angle between $\vec{B}_{\text{SO}}$ and $\hat{g} \cdot \vec{B}$; here, $\hat{g}$ is the $g$-tensor. The $g$-factor in Ge HW has been demonstrated to be extremely anisotropic, with $g_\perp > g_\parallel$[73], leading to a much smaller $B_{\text{C}}$ along the $z$-axis, which is consistent with our experimental result.

To systematically study the anisotropy spectroscopy and determine the SOF direction, a fixed magnetic field is rotated in the $x$-$y$, $x$-$z$, and $y$-$z$ planes while the leakage is measured in the PSB region. Figure 4 shows the leakage current as a function of detuning energy $\varepsilon$ and rotation angle in three orthogonal planes with a fixed magnetic field of 0.8 T, where spin-flip cotunneling is completely suppressed and the SOI dominates. Combining eq (1) and eq (2), the leakage current $I_{\text{leak}}$ induced by the SOI can be expressed by

$$I_{\text{leak}} = I_{\text{SO}}^0 \frac{B^2}{B^2 + \left(\frac{B_{\text{C}}^0}{g^* \cdot \sin\alpha}\right)^2} + I_{\text{B}}'. \tag{3}$$

Based on the $g$-factor in the plane we obtained above, the $g$-tensor for holes in our DQD is well approximated by

$$\hat{g} \approx \begin{pmatrix} 1.2 & 0 & 0 \\ 0 & 1.2 & 0 \\ 0 & 0 & g_z \end{pmatrix}, \tag{4}$$

where $g_z$ is the $g$-factor along the $z$-direction. Considering the effect of SOI in our system, the Rashba SOI will produce a correction to the in-plane $g$-factor and the Dresselhaus SOI will lead to an off-diagonal element $g_{xy}$, which is proportional to the spread of the wave function in the $z$-direction[74]. Because of the strong out-of-plane confinement in our system (the height of Ge HW is around 2 nm), we expect that the value of $g_{xy}$ should be less than 0.1, which is much smaller than the larger diagonal elements and take it as zero for simplicity. Hence, the effective $g$-factor along the external magnetic field can be expressed by

$$g^*(\varphi, \theta) = \sqrt{\hat{r}^\dagger(\varphi, \theta) \hat{g}^\dagger \hat{g} \hat{r}(\varphi, \theta)}, \tag{5}$$

where

$$\hat{r}^{\dagger}(\varphi, \theta) = (\sin\theta\cos\varphi, \sin\theta\sin\varphi, \cos\theta) \quad (6)$$

is a unit vector pointing along the magnetic field direction. The angle $\alpha$ between $\vec{B}_{SO}$ and $\hat{g} \cdot \vec{B}$ can also be expressed based on the $g$-tensor $\hat{g}$ and the direction of $\vec{B}$, $\vec{B}_{SO}$.

Figures 4d-4f show the leakage current measured at $\varepsilon = 0$, which changes periodically with the angle of the magnetic field. From the theoretical expression of eq (3), the extreme anisotropy in the leakage current induced by the SOI in Figure 4 can be explained by two different processes: when the magnetic field is rotated in the *x-z*, *y-z* plane, the experimental results in Figures 4a and 4b are caused by the high anisotropy of $g$-factor in Ge HWs[73]. In our system, the effective $g$-factor reaches its maximum (minimum) when the field points perpendicular (parallel) to the substrate of the nanowires, leading to a minimum (maximum) of $B_C$ and a maximum (minimum) of $I_{\text{leak}}$, as shown in eq (3), which is consistent with our experimental results in Figures 4d and 4e. When the magnetic field is rotated in the *x-y* plane (Figure 4c), the angle between $\vec{B}$ and $\vec{B}_{SO}$ changes, which affects the degree of hybridization of T(1,1) and S(1,1)[39-41]. Especially when the angle between $\vec{B}$ and $\vec{B}_{SO}$ reaches a minimum (i.e., $\vec{B}$ is aligned with the projection of $\vec{B}_{SO}$ in-plane), the spin-flip tunnelling induced by the SOI is suppressed the most, leading to a minimum leakage current in Figure 4f.

To accurately determine the direction of the SOF $\vec{B}_{SO}$, we simultaneously perform numerical fitting of the leakage current measured in three orthogonal planes with eq (3). The red solid lines in Figures 4d-4f represent the theoretical fits according to eq (3) and show great agreement with our experimental results. From these fits, we can extract the SOF direction, which corresponds to $(\theta_0, \varphi_0) = (90° \pm 10°, 31° \pm 5°)$, and the value of $g_z = 3.9 \pm 0.1$. The $g$-tensor we extracted is consistent with ones obtained in the literature before[73]. The ground state in Ge HWs is expected to be an almost pure heavy-hole (HH) state with a very small light-hole (LH) admixture (less than 1%). However, the PSB we studied here is in the region of multiple holes in the QD. This finite occupation should increase the HH-LH mixing and lead to an increase in $g_\parallel$ and a decrease in $g_\perp$, which agrees well with the experimental results in ref 73. Compared to other one-dimensional systems such as Ge/Si core/shell nanowires, a significant feature of the Ge HWs studied here is their triangular cross-section[53]. The lack of inversion symmetry of the cross-section can lead to a large intrinsic SOI that is dependent on the orientation of the wire even without external electric fields[75]. Because the wires grow horizontally along either the [100] or [010] direction[53], the intrinsic SOF points in the *x*-direction according to the theoretical model in ref 75. Notably, the interface inversion asymmetry can also contribute to the SOI[2,76-78]. A Dresselhaus SOI may exist in our system due to the anisotropy of the chemical bonds at material interfaces[76]. Including the spin-orbit terms we mentioned above, the DRSOI, interface SOI and intrinsic SOI induced by the asymmetry of the cross-section, the SOI Hamiltonian in our system can be described by

$$H_{SO} = \alpha k_y \sigma_x + \beta k_y \sigma_y. \tag{7}$$

Here, $\alpha$ is the Rashba interaction coefficient, which accounts for the effect of the DRSOI and intrinsic SOI, $\beta$ is the Dresselhaus interaction coefficient, which corresponds to the interface SOI, $k_y$ is the hole wave operator along the wire, and $\sigma_x$ and $\sigma_y$ are the Pauli spin matrices. Hence, the direction of the SOF in our system can be expressed as $(\alpha, \beta, 0)$, and we can extract the ratio between the Rashba and Dresselhaus SOI in our system $\beta/\alpha = \tan 31° \approx 0.6$. Because the DRSOI is related to the electric field[55,56], the direction of the SOF $\vec{B}_{SO}$ can be adjusted by the gate voltages we apply. The existence of Dresselhaus SOI in our system makes it possible to find operational sweet spots[75] where decoherence effect from charge noise is reduced. By controlling the growth of nanowires along different crystal directions, we can completely turn off SOI when two types of SOI are of equal strength but with opposite directions, realizing the persistent spin helix symmetry[3].

In conclusion, we have measured the leakage current through a DQD in Ge HWs in the PSB regime. We observe and distinguish two different PSB lifting mechanisms:

spin-flip cotunneling and SOI, which lead to an increase in the leakage current. When rotating the direction of magnetic field, we observe the high anisotropy of the leakage current induced by the SOI and explain the experimental results with a combination of the anisotropic hole $g$-factor and the angle between $\vec{B}_{\text{SO}}$ and $\hat{g} \cdot \vec{B}$. By analysing the behavior of the leakage current anisotropy, the direction of the SOF $\vec{B}_{\text{SO}}$ is determined and found to be in-plane with an azimuthal angle of 59° to the nanowire, indicating a large interface SOI along the nanowire. The results we obtained may inspire future developments in the operation of spin qubits and the detection of Majorana fermions.

## Acknowledgements

This work was supported by the National Key Research and Development Program of China (Grant No.2016YFA0301700), the National Natural Science Foundation of China (Grants No. 12074368, 61674132, 12034018, 11625419 and 61922074), the Strategic Priority Research Program of the CAS (Grant No. XDB24030601), the Anhui initiative in Quantum Information Technologies (Grants No. AHY080000), X. H. acknowledge financial support by U.S. ARO through No. W911NF1710257, and this work was partially carried out at the USTC Center for Micro and Nanoscale Research and Fabrication.

wells: from ideal Si membranes to realistic heterostructures. *New Journal of Physics.* **2011**, *13*, 013009.

**Figure Captions:**

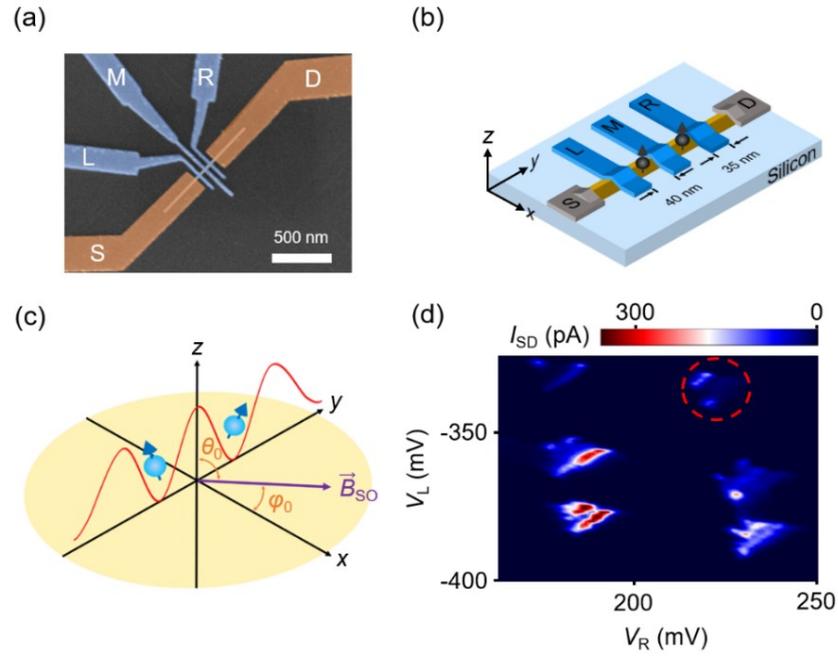

**Figure 1.** (a) False-coloured scanning electron microscopy (SEM) image of a hole DQD in a Ge HW. (b) A 3D schematic of the device. The DQD is formed between two top gates by applying voltages to three grid electrodes. (c) Schematic of the confinement potential created along the nanowire by a gate voltage and the formation of a DQD. Definition of the angles and presentation of the vector for the SOF. (d) Charge stability diagram of the DQD: transport current $I_{SD}$ measured as a function of $V_L$ and $V_R$ with $V_{SD} = 2.5$ mV. The experiment is performed in the bias triangle circled in red.

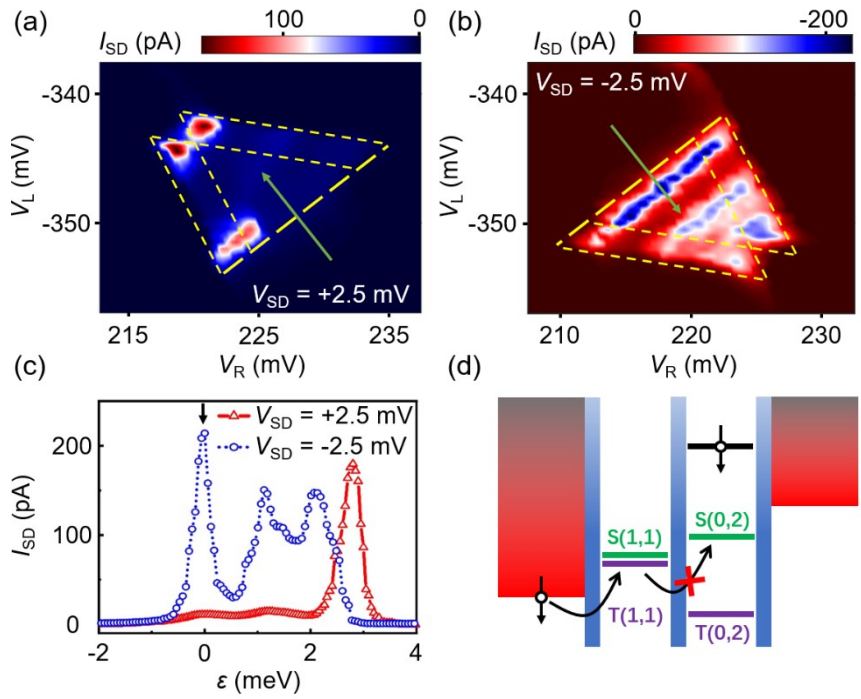

**Figure 2.** (a-b) The zoom-ins of the bias triangle circled in red in Figure 1d with opposite biases. The suppression of current $I_{SD}$ at the base of the bias triangle in panel a is the signature of PSB. The green arrow indicates the direction of the detuning energy $\varepsilon$. (c) The current is measured along the detuning energy with opposite biases (d) Schematic diagram of PSB in a hole DQD. The transport between states (1,1) and (0,2) is forbidden when T(1,1) is occupied.

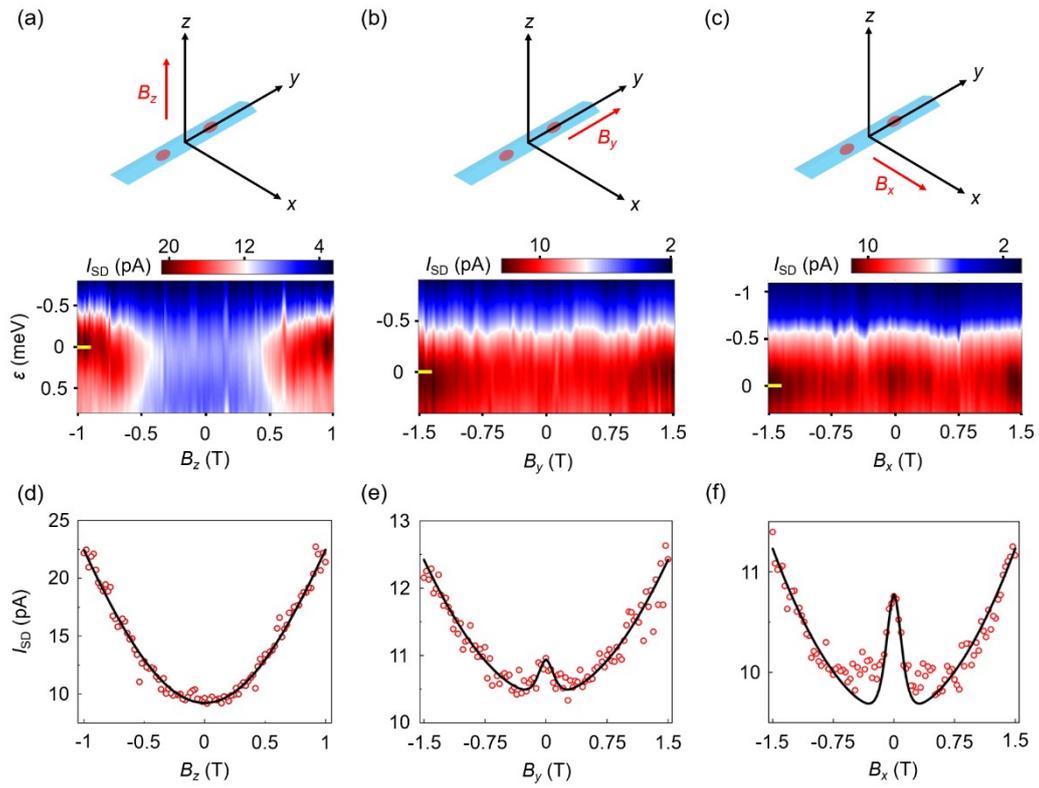

**Figure 3.** (a-c) The measurement of leakage spectroscopy as a function of $\varepsilon$ and $\vec{B}$ in different directions. The direction of magnetic field is depicted in the top panels. The Ge HW is along the y-axis. (d-f) The magnetic field dependence of leakage current along three different directions at $\varepsilon = 0$ (indicated by the yellow bars). The black solid lines represent the theoretical fits of eq (1).

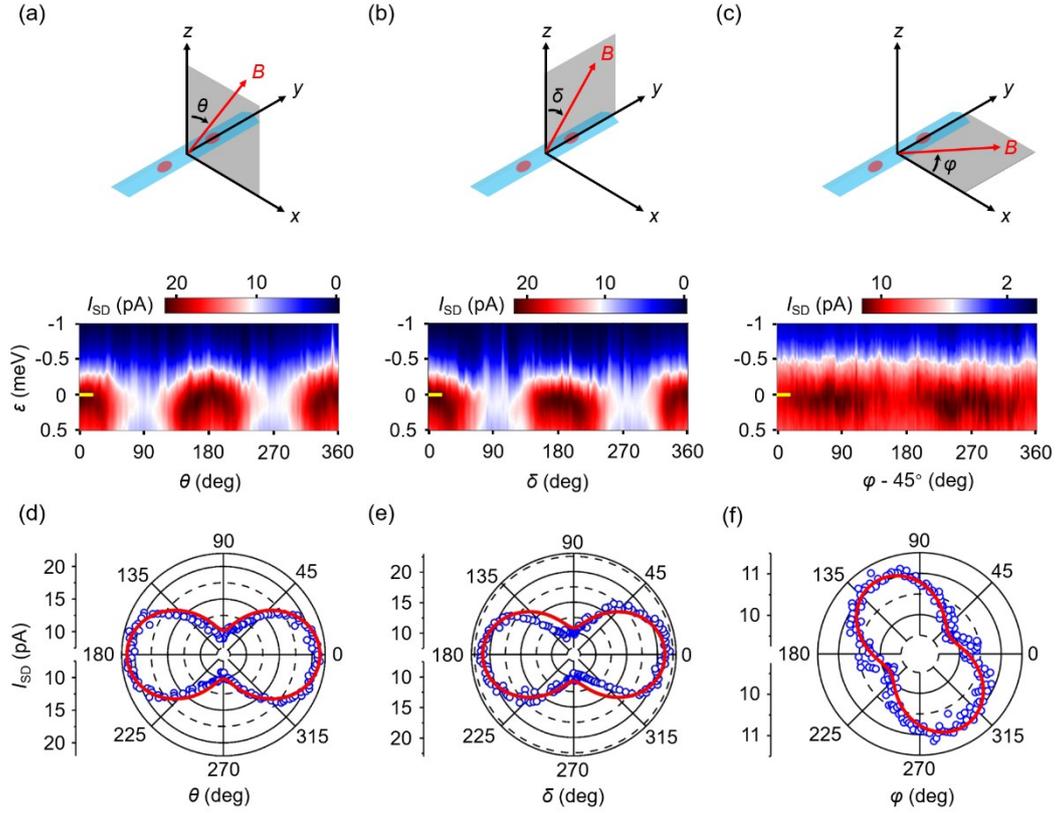

**Figure 4.** (a-c) The anisotropy spectroscopy of leakage current $I_{SD}$ as a function of the detuning energy $\varepsilon$ and the direction of the magnetic field which is rotated in three orthogonal planes as illustrated in the top panel. (d-f) The leakage current as a function of the magnetic field angle at $\varepsilon = 0$ (indicated by the yellow bars) in three different planes. The red solid lines represent the theoretical fits of eq (3).

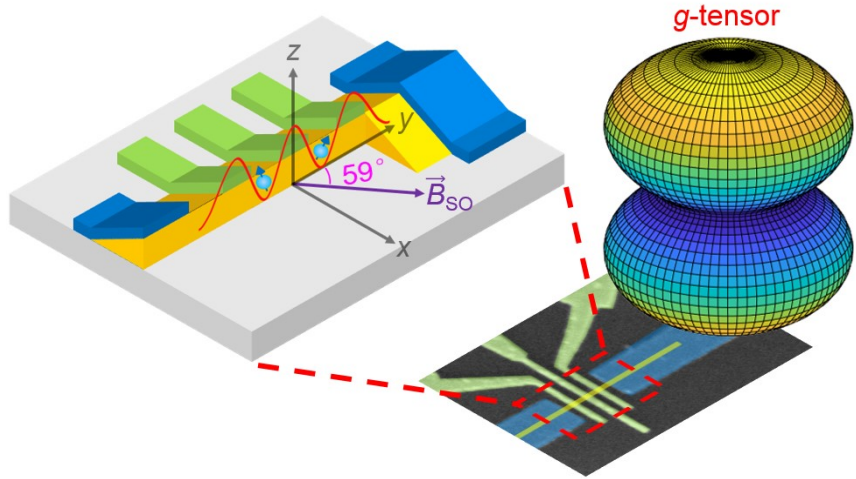

TOC